\documentclass[twocolumn,showpacs,amsfonts,aps,prc,nofootinbib,floatfix,superscriptaddress]{revtex4}

\usepackage{mathrsfs}
\usepackage{epsfig}
\usepackage{amssymb}
\usepackage{amsfonts}
\usepackage{latexsym}
\usepackage{amsthm}

\usepackage{bm}
\usepackage{graphicx}
\usepackage{amsmath}
\usepackage{hyperref}
\usepackage{slashed}

\def\la{\langle}
\def\ra{\rangle}

\begin{document}


\title{Shape and flow fluctuations in ultra-central Pb+Pb collisions at the LHC}

\author{Chun Shen}
\email[Correspond to\ ]{chunshen@physics.mcgill.ca}
\affiliation{Department of Physics, The Ohio State University,
  Columbus, Ohio 43210-1117, USA}
\affiliation{Department of Physics, McGill University, 3600 University Street, Montreal, QC, H3A 2T8, Canada}
\author{Zhi Qiu}
\affiliation{Department of Physics, The Ohio State University,
  Columbus, Ohio 43210-1117, USA}
\author{Ulrich Heinz}
\affiliation{Department of Physics, The Ohio State University,
  Columbus, Ohio 43210-1117, USA}
  
\begin{abstract}
In ultra-central heavy-ion collisions, anisotropic hydrodynamic flow is generated by density fluctuations in the initial state rather than by geometric overlap effects. For a given centrality class, the initial fluctuation spectrum is sensitive to the method chosen for binning the events into centrality classes. We show that sorting events by total {\it initial} entropy or by total {\it final} multiplicity yields event classes with equivalent statistical fluctuation properties, in spite of viscous entropy production during the fireball evolution. With this initial entropy-based centrality definition we generate several classes of ultra-central Pb+Pb collisions at LHC energies and evolve the events using viscous hydrodynamics with non-zero shear but vanishing bulk viscosity. Comparing the predicted anisotropic flow coefficients for charged hadrons with CMS data we find that both the Monte Carlo Glauber (MC-Glb) and Monte Carlo Kharzeev-Levin-Nardi (MC-KLN) models produce initial fluctuation spectra that are incompatible with the measured final anisotropic flow power spectrum, for any choice of the specific shear viscosity. In spite of this failure, we show that the hydrodynamic model can qualitatively explain, in terms of event-by-event fluctuations of the anisotropic flow coefficients and flow angles, the breaking of flow factorization for elliptic, triangular and quadrangular flow measured by the CMS experiment. For elliptic flow, this factorization breaking is large in ultra-central collisions. We conclude that the bulk of the experimentally observed flow factorization breaking effects are qualitatively explained by hydrodynamic evolution of initial-state fluctuations, but that their quantitative description requires a better understanding of the initial fluctuation spectrum. 
\end{abstract}


\date{\today}

\maketitle

\section{Introduction}
\label{sec1}

Heavy-ion collisions at the Relativistic Heavy-Ion Collider (RHIC) and the Large Hadron Collider (LHC) offer a unique window to study the physics of strongly interacting matter in the quark-gluon plasma (QGP) phase. An interesting fundamental property of the QGP is its viscosity which, in combination with the expansion rate of the fireball created in the heavy-ion collision, controls its fluidity. Specifically, the shear viscosity to entropy density ratio $\eta/s$ controls the efficiency with which spatial inhomogeneities and anisotropies of the initial pressure gradients in the fireball are converted to anisotropies in the final collective flow pattern \cite{Heinz:2013th}. Spatial anisotropies in the initial pressure distribution (related to the initial energy density profile through the equation of state (EOS)) are typically characterized in terms of a set of initial eccentricity coefficients $\varepsilon_n$ (defined below as Fourier coefficients of the initial energy density distribution in the plane transverse to the beam direction), while the final flow anisotropies are quantified through coefficients $v_n$ obtained from a Fourier decomposition of the final transverse momentum distribution of the emitted hadrons. Large shear viscosity tends to smooth out flow anisotropies and thus suppress the conversion efficiency $v_n/\varepsilon_n$ \cite{Alver:2010dn,Schenke:2011bn}. 

The experimental observation of large flow anisotropies at both RHIC \cite{Muller:2006ee} and LHC \cite{Muller:2012zq} indicates that the QGP has very low specific shear viscosity $\eta/s$ \cite{Heinz:2013th} close to its quantum limit \cite{Danielewicz:1984ww,Policastro:2001yc}. However, since the extraction of $\eta/s$ from experimental data goes through the conversion efficiency $v_n/\varepsilon_n$ \cite{Song:2010mg}, it requires, in addition to a measurement of $v_n$, a knowledge of the initial eccentricity spectrum $\varepsilon_n$. As the initial density profile of the fireball created in the collision fluctuates from event to event, so do its initial eccentricity and final flow power spectra, $\varepsilon_n$ and $v_n$. At finite impact parameter the eccentricity spectrum $\varepsilon_n$ receives contributions from both collision geometry and event-by-event density fluctuations. In ultra-central collisions, on the other hand, overlap geometry and associated model uncertainties can be largely eliminated as a contributing factor \cite{Luzum:2012wu}, focussing attention on the initial spectrum of quantum fluctuations of the strongly interacting quantum fields within the colliding nuclear wave functions. These are notoriously difficult to compute theoretically, making experimental studies of ultra-central heavy-ion collisions \cite{ALICE:2011ab,ATLAS:2012at,CMS:2013bza} particularly valuable for improving our quantitative understanding of heavy-ion collision dynamics.

\begin{figure*}[ht!]
  \centering
  \includegraphics[width=0.83\linewidth]{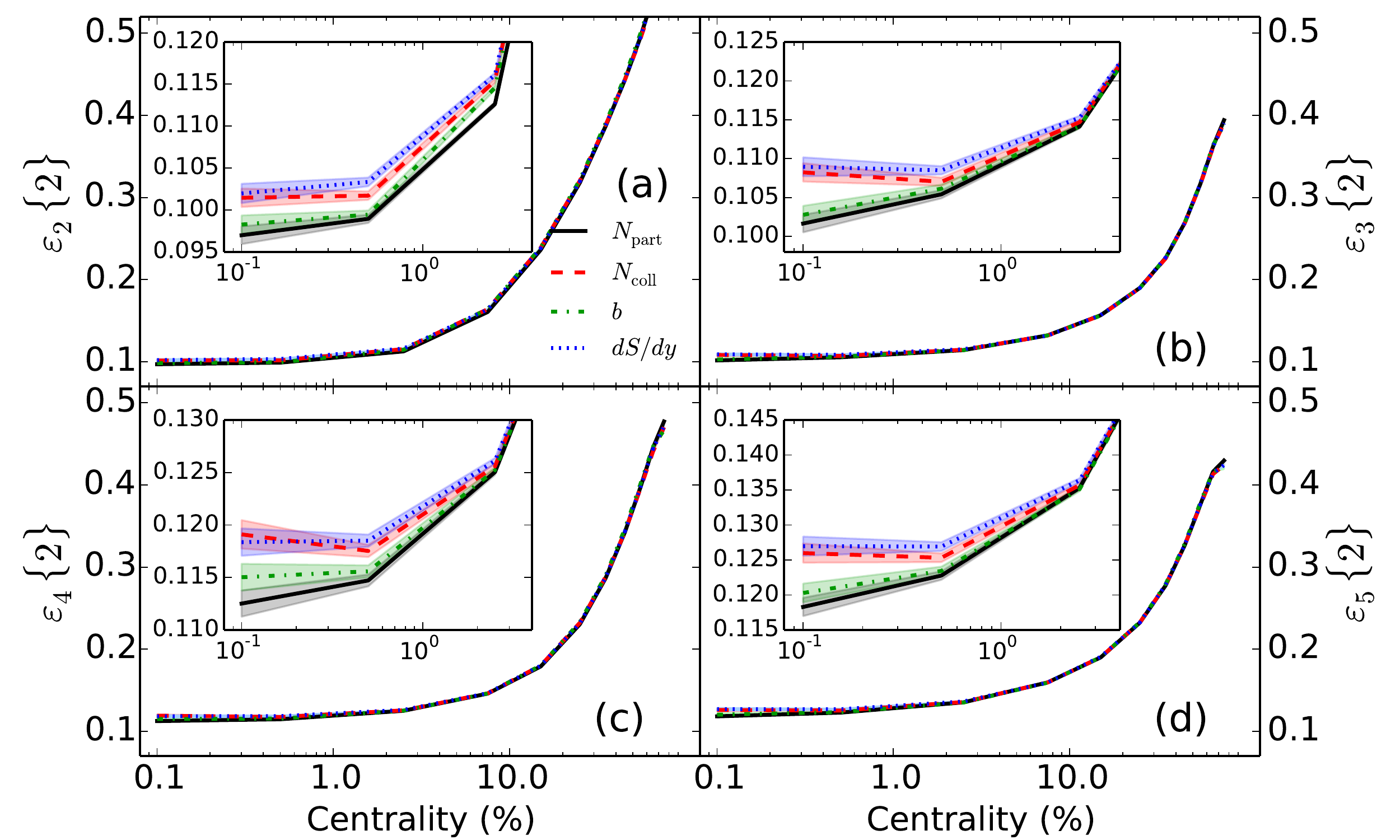}
  \caption{(Color online) The root mean square (rms) initial eccentricity $\epsilon_n\{2\}$ for $n{\,=\,}2{-}5$, as a function of centrality, for different centrality selection criteria: using the number of participant nucleons $N_\mathrm{part}$ (solid black), the number of binary collisions $N_\mathrm{coll}$ (red dashed), impact parameter $b$ (green dotted), or initial total entropy $dS/dy$ (dotted blue). For this figure 1 million events were generated from the MC-Glauber model with minimum inter-nucleon distance (``hard core radius'') $r_\mathrm{min}=0.9$\,fm \cite{Broniowski:2010jd,Luzum:2013yya}, for Pb+Pb collisions at $\sqrt{s}=2.76\, A$\,TeV, with binary collision to wounded nucleon ratio $\alpha = 0.118$. }
  \label{fig1}
\end{figure*}

It bears pointing out that a measurement of the final flow power spectrum $v_n$ in ultra-central collisions between equal-size nuclei provides access to both this initial fluctuation spectrum \cite{Mishra:2007tw,Mocsy:2011xx} (the ``Little Bang temperature power spectrum'' \cite{Heinz:2013wva}) and the QGP transport coefficients, especially its shear viscosity \cite{Heinz:2013th,Staig:2010pn,Lacey:2013is}. Since the specific shear viscosity $\eta/s$ controls the anisotropic flow response of the expanding liquid to the initial density fluctuations, and neither the specific shear viscosity nor the initial fluctuation spectrum are theoretically known with any precision, a phenomenological analysis of anisotropic flow fluctuations must be sufficiently comprehensive to be able to constrain both simultaneously.

The CMS Collaborations at the LHC has measured \cite{CMS:2013bza} the anisotropic flow coefficients of charged hadrons in ultra-central Pb+Pb collisions at $\sqrt{s} = 2.76$\,$A$\,TeV, using events from the tail of the charged multiplicity distribution corresponding to $0{-}0.2\%$ centrality. Implementing such tight multiplicity cuts is a challenge for theoretical simulations which have typically much smaller event samples than the experimental measurements. In Sec.~\ref{sec2} of this work, we investigate how different ways of cutting centrality bins can affect the initial-state eccentricity spectrum, and show how to faithfully mimic the experimental centrality selection procedure without the need for evolving astronomically large numbers of initial configurations. In Sec.~\ref{sec3}, we show the charged hadron particle spectrum and its anisotropic flow coefficients. In Sec.~\ref{sec4} we check the flow factorization ratios in ultra-central Pb+Pb collisions at the LHC. We summarize our findings in Sec.~\ref{sec5}. 

\section{Initial state fluctuations and event selection in ultra-central nucleus-nucleus collisions}
\label{sec2}

Experimentally, the ultra-central collision events are defined as those with the highest measured final charged multiplicities. However, in theoretical calculations, using final charged multiplicities to determine the event centrality is numerically very expensive. Especially for the narrow 0-0.2\% centrality bin, the numerical effort spent on evolving the remaining 99.8\% of the simulated events is wasted. So we would like to determine (at least estimate) the event centrality of each initial profile before evolving them through viscous hydrodynamics. 

\begin{table*}
\begin{center}
\begin{tabular}{|c||c|c||c|c||c|c||c|c|}
\hline
 & (0-0.2\%)$_{dN/dy}$ & (0-0.2\%)$_{dS/dy}$ & (0-1\%)$_{dN/dy}$ & (0-1\%)$_{dS/dy}$ & (1-2\%)$_{dN/dy}$ & (1-2\%)$_{dS/dy}$ & (2-3\%)$_{dN/dy}$ & (2-3\%)$_{dS/dy}$ \\ \hline \hline 
 $\varepsilon_2\{2\}$ &  $0.082 \pm 0.002$ & $0.081 \pm 0.002$ & $0.085 \pm 0.001$ & $0.086 \pm 0.001$ & $0.096 \pm 0.001$ & $0.096 \pm 0.001$ & $0.114 \pm 0.001$ & $0.114 \pm 0.001$ \\ \hline 
 $\varepsilon_3\{2\}$ & $0.084 \pm 0.002$ & $0.085 \pm 0.002$ & $0.088 \pm 0.001$  &  $0.088 \pm 0.001$ & $0.092 \pm 0.001$  &  $ 0.092 \pm 0.001$ & $0.097 \pm 0.001$  & $ 0.096 \pm 0.001$ \\ \hline
 $\varepsilon_4\{2\}$ & $0.095 \pm 0.002$  &  $0.096 \pm 0.002$ & $0.095 \pm 0.001$  &  $0.095 \pm 0.001$ & $0.097 \pm 0.001$  &  $ 0.097 \pm 0.001$ & $0.102 \pm 0.001$ &  $ 0.103 \pm 0.001$ \\ \hline 
 $\varepsilon_5\{2\}$ & $0.089 \pm 0.002$  &  $0.092 \pm 0.002$ & $0.097 \pm 0.001$  &  $0.097 \pm 0.001$ &  $0.104 \pm 0.001$ &  $ 0.104 \pm 0.001$ & $0.110 \pm 0.001$ &  $ 0.110 \pm 0.001$ \\ \hline
 \end{tabular}
 \end{center}
\caption{Comparison of the initial eccentricities $\varepsilon_n\{2\}$ ($n{\,=\,}2{-}5$), computed with the MC-KLN model with a minimum distance (``hard core'') $r_\mathrm{min}=0.4$\,fm between nucleons \cite{Hirano:2009ah,Qiu:2012uy}, in central and ultra-central centrality bins that are determined by initial total entropy or final charged hadron multiplicity, respectively.}
\label{table0}
\end{table*}

We begin by showing that for ultra-central collisions the initial fluctuation spectrum $\{\varepsilon_n\{2\}|n{\,=\,}1,2,\dots\}$ (whose precise definition is given further below) can depend strongly on the method for determining centrality. We study four popular choices of the collision parameters ($N_\mathrm{part}$, $N_\mathrm{coll}$, $b$, $dS/dy$) to categorize the event centrality of every initial density profile generated from the MC-Glauber model \cite{Shen:2014vra}. The produced initial entropy density at $\tau_0$ in the transverse plane is calculated as
\begin{equation}
s({\bf r_\perp}; \tau_0) = \frac{\kappa}{\tau_0} 
\left( \frac{1{-}\alpha}{2}\, n_\mathrm{WN}(\bm{r}_\perp) 
        + \alpha\, n_\mathrm{BC}(\bm{r}_\perp)\right),
\label{eq0.0}
\end{equation}
where $\kappa$ is an overall normalization factor, tuned to reproduce the final charged hadron multiplicity $dN^\mathrm{ch}/d\eta$ at mid-rapidity, and $\alpha$ is the mixing ratio between the wounded nucleon ($n_\mathrm{WN}(\bm{r})$) and binary collision ($n_\mathrm{BC}(\bm{r})$) profiles \cite{Shen:2014vra}. We choose $\alpha = 0.118$ \cite{Qiu:2011hf} to reproduce the measured centrality dependence of $dN^\mathrm{ch}/d\eta$ in 2.76\,$A$\,TeV Pb+Pb collisions. 

In Fig.~\ref{fig1}, the initial eccentricities $\varepsilon_n\{2\}$ ($n{\,=\,}2{-}5$) are shown as functions of centrality. For semi-peripheral and peripheral collisions, the $\varepsilon_n\{2\}$ spectrum is insensitive to the method used to determine the centrality bins. However, differences become noticeable when we select events within the top 1\% centrality range. We notice that using the initial total entropy to determine the centrality bin gives the largest initial eccentricities $\{\varepsilon_n\{2\}\}$, about 10\% larger than if we use the number of participant nucleons. So for a narrow bin of very central collisions, extra care is needed in the event selection. 

\begin{figure}[b!]
  \centering
  \includegraphics[width=0.95\linewidth]{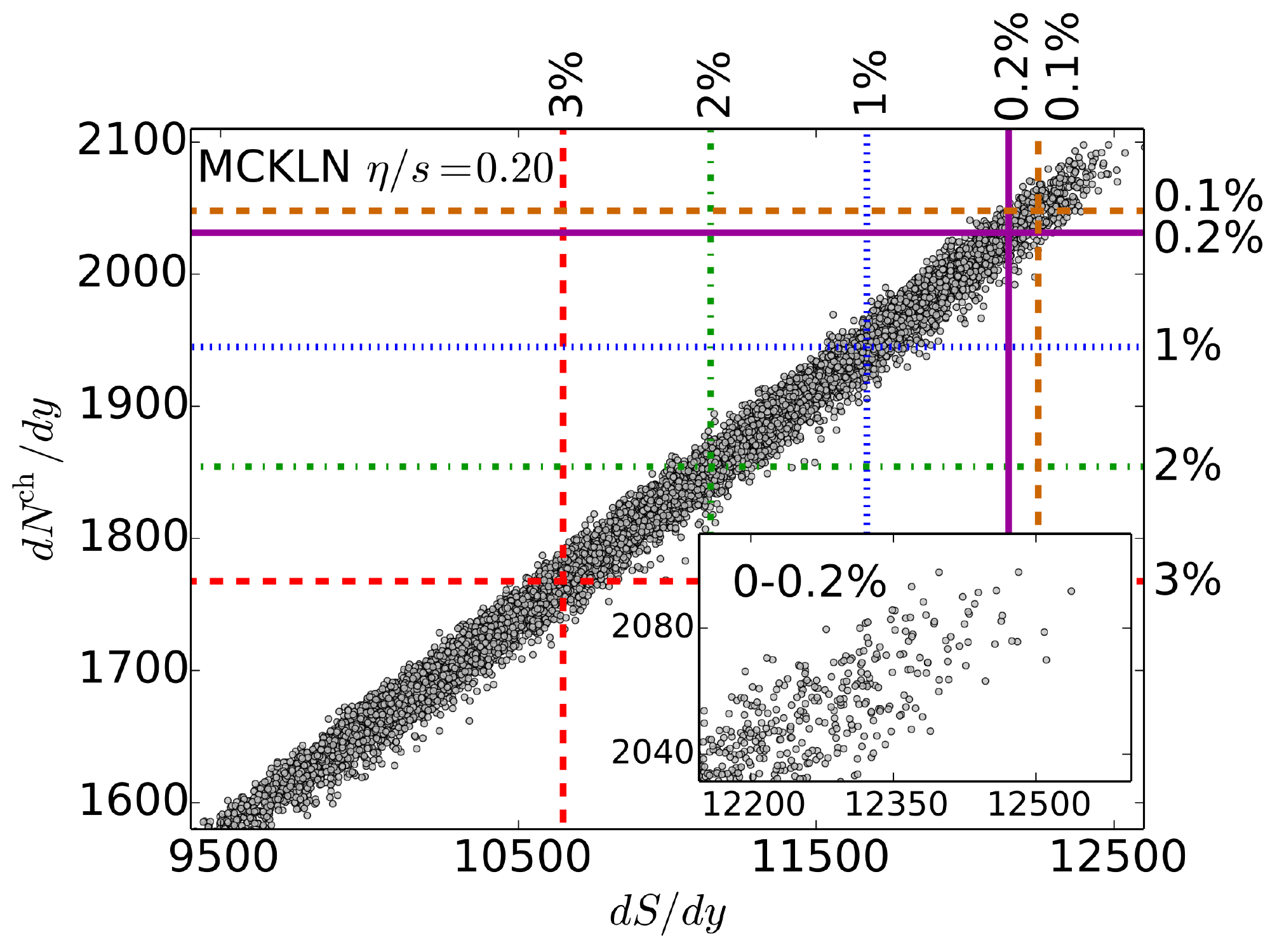}
  \caption{(Color online) Correlation between the initial total entropy $dS/dy$ and final charged multiplicity
  $dN_\mathrm{ch}/dy$ for (ultra-)central Pb+Pb collisions at the LHC (0-5\% centrality). The 
  plot is based on 10,000 previously generated viscous hydrodynamic events with fluctuating MC-KLN 
  initial conditions using a minimum inter-nucleon distance $r_\mathrm{min}=0.4$\,fm 
  \cite{Hirano:2009ah,Qiu:2012uy} and evolved with $\eta/s{\,=\,}0.2$. The lines separate different 
  centrality classes (as indicated), defined by ordering the events according to $dN_\mathrm{ch}/dy$ 
  (horizontal lines) or $dS/dy$ (vertical lines). The normalized variance of the distribution is 
  about 0.6\% in both vertical and horizontal directions.}
  \label{F2}
\end{figure}

\begin{figure}[h!]
  \centering
  \includegraphics[width=0.95\linewidth]{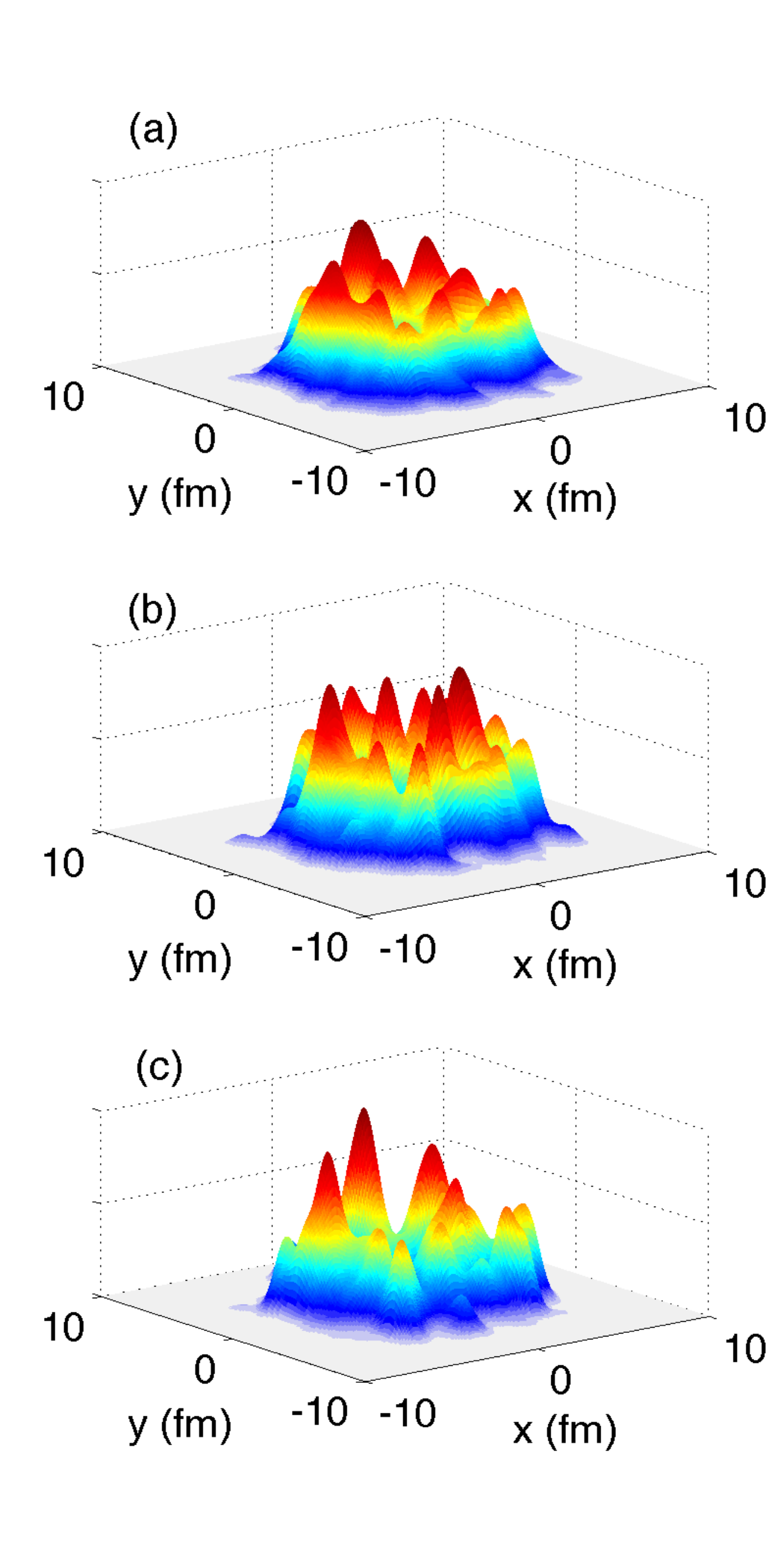}
  \caption{(Color online) Initial entropy density profiles in the transverse plane from the MC-Glauber
  model, for one selected sampling of the nucleon positions inside the colliding nuclei. Panel (a) does 
  not contain $pp$ multiplicity fluctuations. Panels (b) and (c) correspond to two different samplings
  of the entropy production associated with each nucleon-nucleon collision when $pp$ multiplicity 
  fluctuations are included as described in the text.}
  \label{F2.1}
\end{figure}

Experience based on the hydrodynamic evolution of smooth, ensemble-averaged initial conditions has shown that the observed charged multiplicity in the final state, $dN_\mathrm{ch}/d\eta$, used by the experiments for cutting on centrality, is monotonically related to the initial entropy density $dS/dy$. Figure~\ref{F2} demonstrates that this remains true on average, within a narrow variance caused by fluctuations of the viscous entropy production in fireballs with particularly rough initial conditions (examples are shown in Fig.~\ref{F2.1}), when one evolves bumpy initial conditions that fluctuate from event to event.%
\footnote{%
    Fig.~\ref{F2} is based on events with MC-KLN initial conditions that were evolved with $\eta/s = 0.20$,
    the largest value for the specific shear viscosity studied in this work. For MC-Glauber initial conditions
    evolved with $\eta/s=0.08$ the viscous entropy production is smaller, and we correspondingly found
    a reduced normalized variance of 0.2\% for the analogous $dS/dy$ vs. $dN_\mathrm{ch}/dy$ 
    distribution. Both sets of hydrodynamic events were previously generated for an earlier study 
    \cite{Qiu:2011hf} and do not include $pp$-multiplicity fluctuations which, as discussed below, tend to
    increase the ``bumpiness'' of the initial density profiles and are therefore expected to somewhat increase
    the normalized variance of the distribution shown in Fig.~\ref{F2}.} 
Note that the events plotted in Fig.~\ref{F2} are all within the $0{-}5\%$ centrality bin where variations in $dS/dy$ are caused almost entirely by fluctuations, and variations in overlap geometry play only a small role.

Table~\ref{table0} shows that, within the statistical uncertainties imposed by our limited number of fully evolved events, the eccentricity coefficients $\varepsilon_{2,3,4,5}$ defined in Eq.~(\ref{eq1}) below for events in a given centrality bin agree for bins defined by ordering the events according to initial $dS/dy$ or final $dN_\mathrm{ch}/dy$. We also checked and confirmed that the  distributions of these coefficients within the bins agree for both centrality definitions. Based on all these observations, we proceeded to generate 1 million minimum bias Pb+Pb events as described next, ordered them by their initial total entropy $dS/dy$, divided them into centrality classes (as shown in Fig.~\ref{F2}) and then took the 0.2\% of them with the largest $dS/dy$ values to represent the 0-0.2\% centrality class of ultra-central Pb+Pb collisions studied by the CMS Collaboration. These events are then used in the rest of the paper. 

\begin{figure}[h!]
  \centering
  \includegraphics[width=0.9\linewidth]{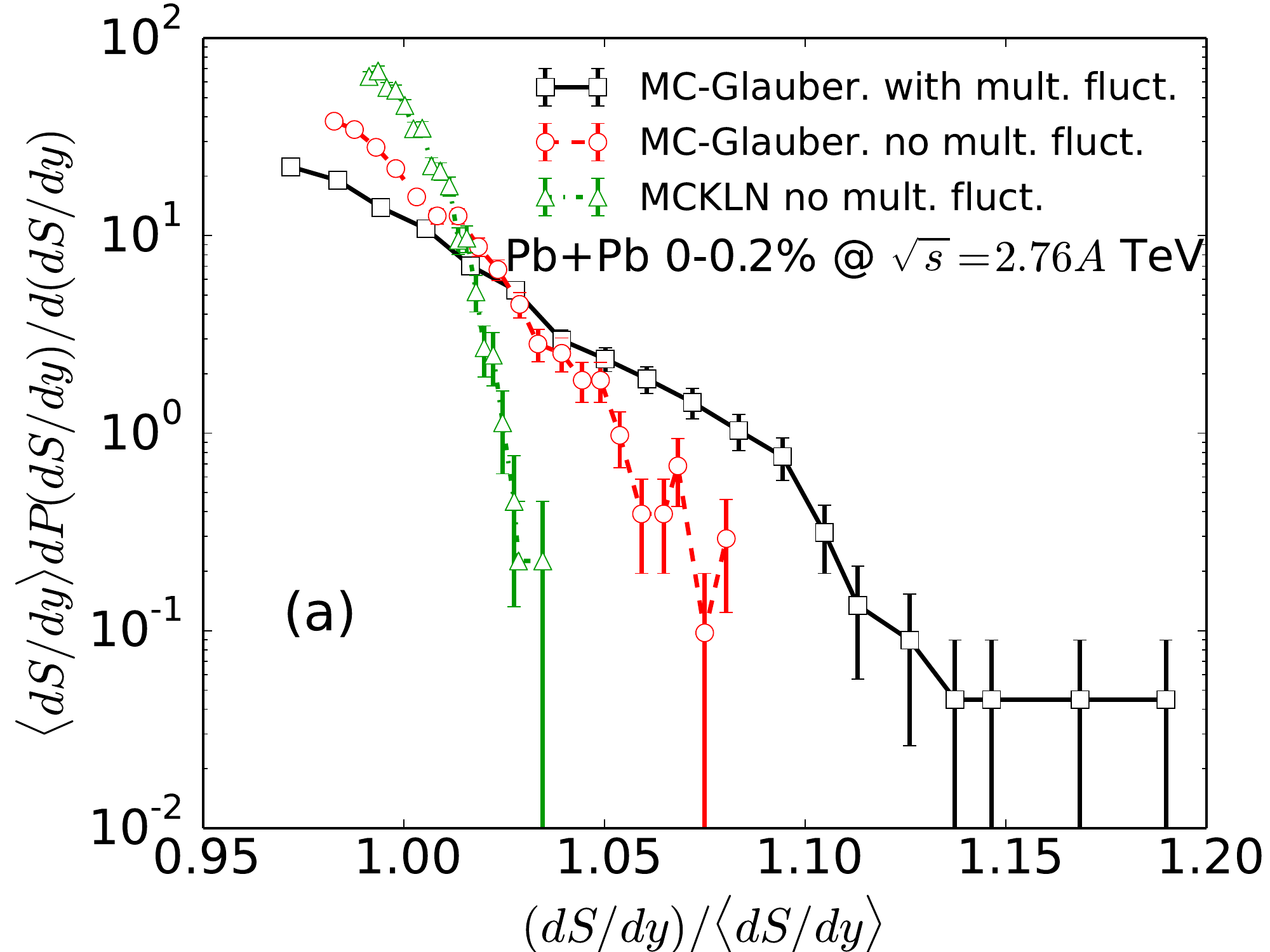}
  \includegraphics[width=0.9\linewidth]{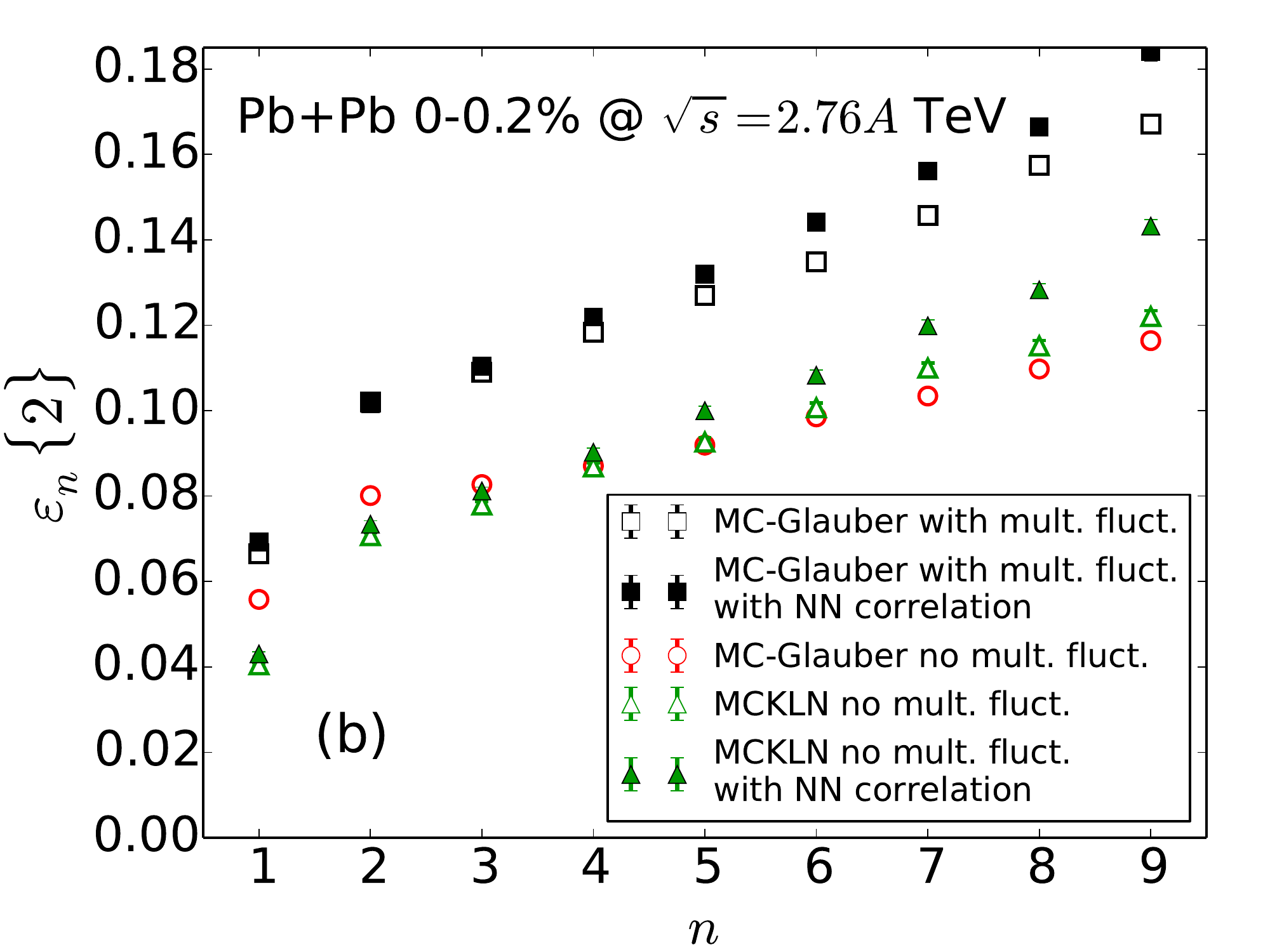}
  \caption{(Color online) The initial total entropy distribution (a) and the initial rms eccentricity 
      $\varepsilon_n \{2\}$ as a function of harmonic order $n$ (b) from the MC-Glauber 
      and MC-KLN models for 0-0.2\% ultra-central Pb+Pb collisions at 2.76\,$A$\,TeV. The 
      results shown with open symbols are calculated requiring a minimum inter-nucleon distance 
      (``hard core radius'') $r_\mathrm{min}=0.9$\,fm \cite{Broniowski:2010jd, Luzum:2013yya,Shen:2014vra}. The filled 
      symbols in panel (b) are obtained by sampling nucleon configurations that incorporate a 
      realistic repulsive 2-body nucleon-nucleon correlation \cite{Alvioli:2009ab}. }
  \label{fig2}
\end{figure}
%

In order to capture the bias in ultra-central collisions towards events whose entropy production fluctuates upward from the mean, we implement $pp$ multiplicity fluctuations in the MC-Glauber model in a way that allows us to reproduce the experimentally observed KNO scaling of the $pp$ multiplicity distributions. Our implementation is described in detail in \cite{Shen:2014vra} (see earlier work in \cite{Broniowski:2007nz,Qin:2010pf,Denicol:2014ywa} for related but slightly different approaches).%
\footnote{%
    For the MC-KLN model we currently do not implement any collision-by-collision multiplicity fluctuations.
    An implementation in the MC-KLN model of KNO scaling in $pp$ coolisions can be found in 
    Ref.~\cite{Dumitru:2012yr}.}
Figure~\ref{F2.1} illustrates the effect of $pp$ multiplicity fluctuations on the shape of initial entropy density profile in Pb+Pb collisions. They tend to increase the variance of the density fluctuations in the initial state, amplifying the magnitude of hot spots and deepening the valleys between them. Overall the initial density distributions become more bumpy without, however, changing the characteristic radius of the hot spots (which is still given by the nucleon size). In Fig.~\ref{fig2}a, we show the initial total entropy distribution for MC-Glauber and MC-KLN initial conditions, for Pb+Pb collisions in the 0-0.2\% centrality bin at LHC energies. Even in the absence of $pp$ multiplicity fluctuations, the MC-Glauber model yields a significantly broader multiplicity distribution in Pb+Pb collisions than the MC-KLN model. By adding $pp$ multiplicity fluctuations to the MC-Glauber model, the chances for upward fluctuations in the multiplicity for ultra-central Pb+Pb collisions are strongly increased. For a correct simulation of ultra-central Pb+Pb initial conditions, proper inclusion of $pp$ multiplicity fluctuations is therefore mandatory. 

The shape fluctuations of the initial conditions can be characterized by a series of complex eccentricity coefficients. For a single event and for $n \ge 2$, the $n$-th order eccentricity of the initial condition is defined as the modulus of the complex quantity
\begin{equation}
\varepsilon_n e^{i n \Phi_n} = - \frac{\int d^2 r\, r^n\, e^{in\phi}\,e(r, \phi)}{\int d^2 r\, r^n\, e(r, \phi)},
\label{eq1}
\end{equation}
in which we use the transverse local rest frame energy density $e(r, \phi)$ as a weight function.  The phase $\Phi_n$ of the complex quantity in Eq.~(\ref{eq1}) defines the $n$-th order participant plane angle. Since the energy density is centered at the origin, Eq.~(\ref{eq1}) gives zero for $n=1$. Hence, one defines $\epsilon_1$ using the next order in the cumulant expansion \cite{Teaney:2010vd}, 
\begin{equation}
\varepsilon_1 e^{i \Phi_1} = - \frac{\int d^2 r\, r^3\, e^{i \phi}\, e(r,\phi)}{\int d^2 r\, r^3\, e(r,\phi)}.
\label{eq1a}
\end{equation}
$\varepsilon_n\{2\}\equiv\sqrt{\langle\varepsilon_n^2\rangle}$ is the root mean square value of these eccentricity coefficients for an ensemble of fluctuating events (where $\langle\dots\rangle$ denotes an ensemble average).

In Fig. \ref{fig2}b we plot the initial fluctuation power spectrum $\varepsilon_n\{2\}$ for the MC-Glauber and MC-KLN models in 0-0.2\% ultra-central collisions at the LHC. At this centrality, the $\varepsilon_n\{2\}$ are entirely dominated by fluctuations in the initial density profile, with no contributions from geometric overlap effects. With the $r^n$ weighting factor in Eq.~(\ref{eq1}), the magnitudes of $\varepsilon_n\{2\}$ are all similar (except for $\varepsilon_1\{2\}$ which is $\sim 40\%$ smaller), for both initial condition models. Without $pp$ fluctuations, the $\varepsilon_n\{2\}$ from the MC-KLN model are close to those from the MC-Glauber model for $n\geq 3$ and about 10-30\% smaller than the MC-Glauber eccentricities for $n{\,=\,}1$ and 2. Inclusion of $pp$ fluctuations in the MC-Glauber model leads to a significant increase of $\varepsilon_n\{2\}$ for all harmonics, ranging from 15\% for $n{\,=\,}1$ to almost 50\% for $n{\,=\,}9$. We also include results (shown by filled symbols) which sample realistic nucleon configurations of a lead nucleus in the presence of repulsive two-nucleon correlations \cite{Alvioli:2009ab}. For both the MC-Glauber and MC-KLN models, realistic 2-body nucleon-nucleon correlations result in low-order eccentricities that are very close to those obtained when simply imposing (as suggested in \cite{Broniowski:2010jd,Luzum:2013yya}) a minimum inter-nucleon distance (``hard core radius'') $r_\mathrm{min} = 0.9$\,fm in the Monte Carlo sampling of the nucleon positions from nuclear density distribution. Higher-order $\varepsilon_n$ with $n{\,\gtrsim\,}6$ are about 10\% larger for samples containing realistic nucleon-nucleon correlations. 

\begin{table}[t!]
\begin{center}
\begin{tabular}{|l||c|c|}
\hline
 Model & $ dN/d\eta \vert_{\vert \eta \vert < 0.5} $ & $\langle p_T \rangle$ (GeV) \\ \hline
MC-Glb. ideal  & 1793.85 $\pm$ 1.02 & 0.715 $\pm$ 0.001\\ \hline
MC-Glb. $\eta/s = 0.08$  & 1799.96 $\pm$ 1.07 & 0.705 $\pm$ 0.001\\ \hline
MC-Glb. $\eta/s = 0.12$  & 1780.44 $\pm$ 1.03 & 0.704 $\pm$ 0.001\\ \hline
MC-Glb. $\eta/s = 0.20$  & 1797.76 $\pm$ 1.68 & 0.711 $\pm$ 0.001\\ \hline
MC-KLN ideal  & 1809.78 $\pm$ 0.60 & 0.688 $\pm$ 0.001\\ \hline
MC-KLN $\eta/s = 0.08$  & 1807.66 $\pm$ 0.43 & 0.682 $\pm$ 0.001\\ \hline
MC-KLN $\eta/s = 0.12$  & 1807.81 $\pm$ 0.43 & 0.685 $\pm$ 0.001\\ \hline
MC-KLN $\eta/s = 0.20$  & 1811.03 $\pm$ 0.53 & 0.692 $\pm$ 0.001\\ \hline
\end{tabular}
\end{center}
\caption{The total yield of charged hadrons and their mean $p_T$ in 0-0.2\% in Pb+Pb collisions at the LHC. In each set of runs we fixed the overall normalization factor to the experimentally measured charged hadron multiplicity at 0-5\% centrality.}
\label{table1}
\end{table}

In the next section, we will evolve these initial conditions using viscous hydrodynamics with four different specific shear viscosities, $\eta/s = 0, 0.08, 0.12, 0.20$. For each set of runs we generate 1000 events which is sufficient for a statistical analysis of anisotropic flows.  

\section{Particle spectra and their flow anisotropies}
\label{sec3}

\begin{figure}[h!]
  \centering
  \includegraphics[width=0.9\linewidth]{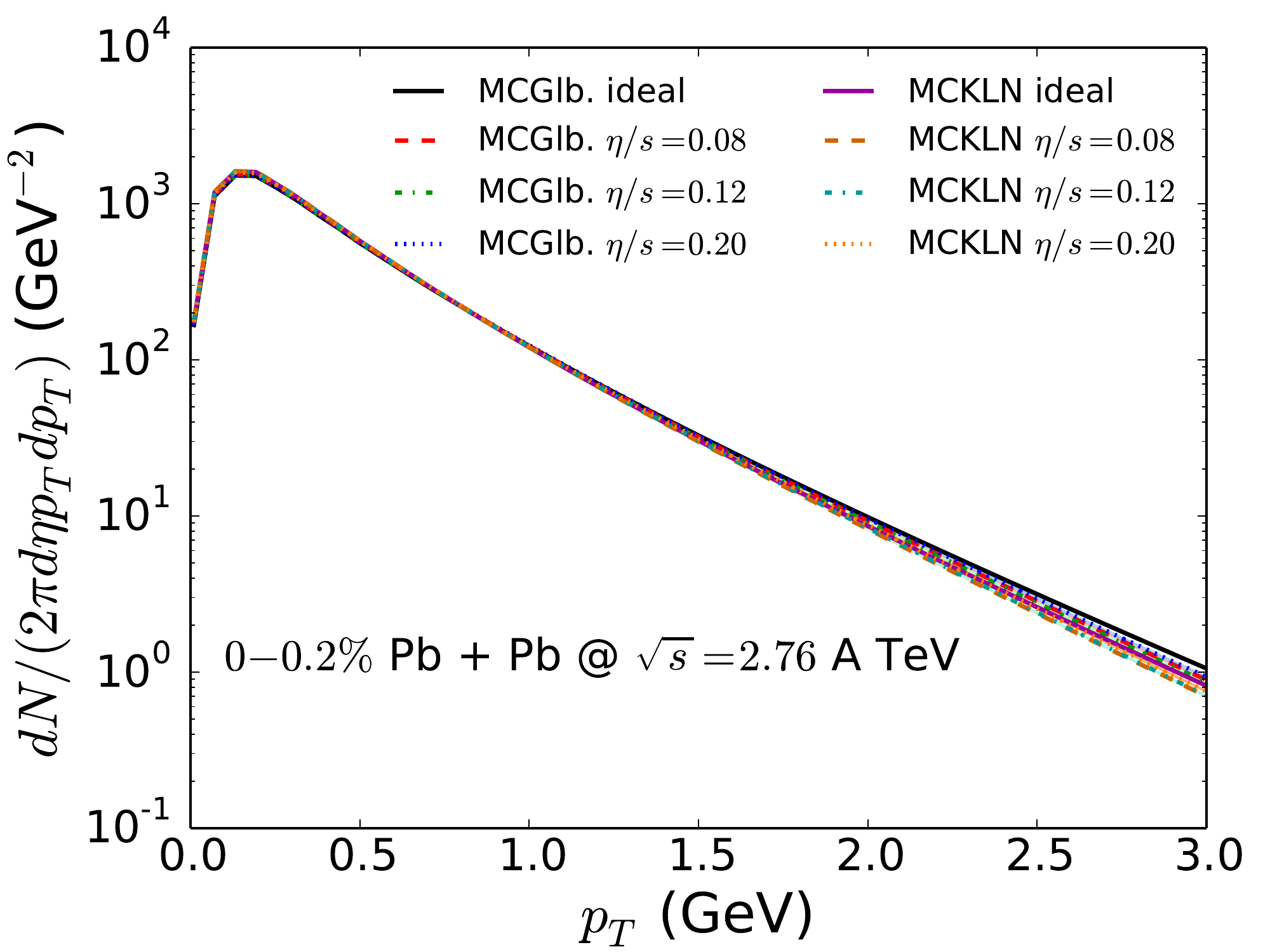}
  \caption{(Color online) Charged hadron $p_T$-spectra from the MC-Glb and MC-KLN models, for different
      specific shear viscosities as indicated, for Pb+Pb collisions of 0-0.2\% centrality
      at $\sqrt{s} = 2.76$\,$A$\,TeV.}
  \label{fig3}
\end{figure}

In Fig.~\ref{fig3} we show the transverse momentum spectra of charged hadrons from the MC-Glauber and the MC-KLN models, hydrodynamically evolved with different $\eta/s$ values. In ultra-central collisions, we find that the shear viscosity has only minor effects on the slope of the charged hadron spectra.

%
%

%
\begin{figure*}
  \centering
  \includegraphics[width=0.85\linewidth]{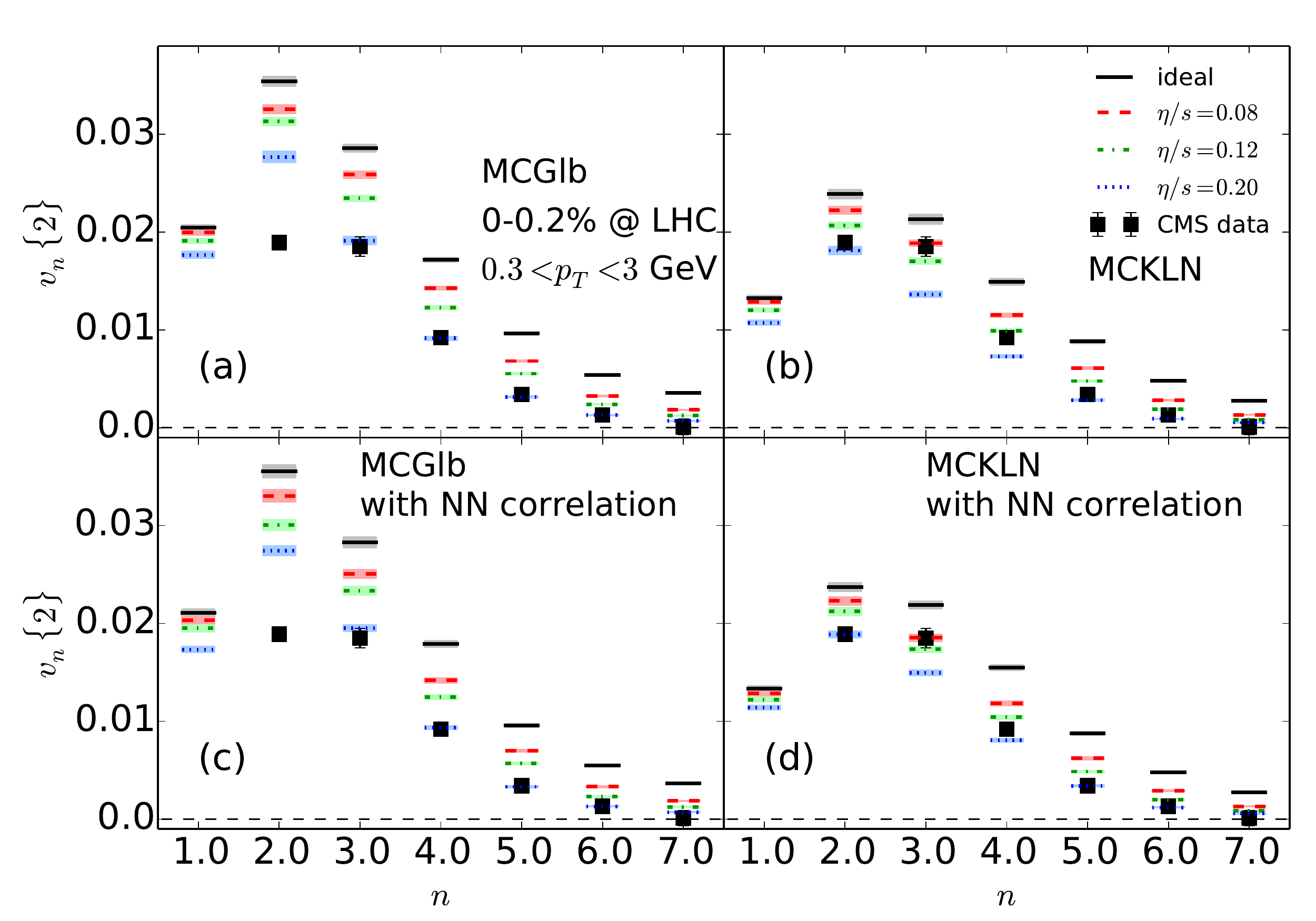}
  \caption{(Color online) $p_T$-integrated $v_n\{2\}$ of charged hadrons in 2.76\,$A$\,TeV 
   Pb+Pb collisions at 0-0.2\% centrality, from the MC-Glauber (a,c) and the MC-KLN 
   models (b,d) models. In (a,b) repulsive hard core correlations among nucleons are
   implemented by simply requiring a minimum inter-nucleon distance 
   $r_\mathrm{min}=0.9$\,fm when sampling the nucleon positions, in (c,d) realistic 
   2-nucleon correlations \cite{Alvioli:2009ab} are included in the sampling routine. 
   The solid squares show the CMS measurements \cite{CMS:2013bza} for comparison. 
    As in the CMS measurements, $v_n\{2\}$ is integrated over $p_T$ from 0.3 to 3 GeV.}
  \label{fig4}
\end{figure*}
%
The MC-Glauber initial conditions, which include $pp$ multiplicity fluctuations, result in slightly flatter (harder) charged particle spectra compared to the MC-KLN model. This is because the additional collision-by-collision multiplicity fluctuations increase the bumpiness of the initial density profiles and thereby the initial pressure gradients, which drives the system to develop more radial flow. In Table \ref{table1} we summarize for the eight cases shown in Fig.~\ref{fig3} the total charged hadron yields and their mean $p_T$ in 0-0.2\% ultra-central Pb+Pb collisions at the LHC. Ideal hydrodynamic simulations result in a larger mean $p_T$ than the viscous ones. Due to the lack of viscous entropy production in an ideal fluid, the system needs to start with a higher peak temperature in order to produce the same final multiplicity. This results a larger pressure gradient and a ${\sim\,}10\%$ longer fireball lifetime, which both lead to the development of stronger radial hydrodynamic flow compared to viscous evolution. Because of the large system size in Pb+Pb collisions, viscous effects on the fireball expansion are not as pronounced as in smaller collision systems. A growth of $\eta/s$ by a factor 2.5 is seen to increase the charged hadron mean $p_T$ by just over 1\%. 

Due to fluctuations, the $p_T$ distributions of individual events are azimuthally anisotropic even for zero impact parameter. These anisotropies can be characterized by a series of ($p_T$ and rapidity dependent) harmonic flow coefficients $v_n$ and associated flow angles $\Psi_n$:
\begin{eqnarray}
\label{eq3}
  && \!\!\!\!\!\!\!\!\!\!\!\!\!\!\!\!  \frac{dN}{dy p_T dp_T d\phi_p} = \frac{dN}{dy p_T dp_T} 
\nonumber\\
  &&\times \biggl( 1 + 2\sum_{n{=}1}^\infty v_n(p_T,y)\cos\bigl(\phi_p{-}\Psi_n(p_T,y)\bigr) \biggr).
\end{eqnarray}
In the absence of non-flow effects, the ``2-particle cumulant flow'' $v_n\{2\}$ is the root mean square of the fluctuating flow $v_n$ in an event ensemble: $v_n\{2\}=\sqrt{\langle v_n^2\rangle}$.
   
In Fig.~\ref{fig4} we show the charged particle $p_T$-integrated $v_n\{2\}$ as a function of its harmonic order $n$ for both the MC-Glauber and the MC-KLN models, for 0-0.2\% ultra-central Pb+Pb collisions at the LHC. The MC-Glauber initial conditions account for $pp$ multiplicity fluctuations. The top and bottom panels compare calculations based on two different approaches for including repulsive 2-nucleon correlations within the nuclei, as described in the figure caption. The differences between them are seen to be small and within the experimental uncertainties of the measured $p_T$-integrated $v_n\{2\}$ values. For the $p_T$-differential flow observables discussed below, the corresponding differences are typically less than the line widths in the plots. In the rest of the paper we therefore show only calculations based on the more realistic modeling of repulsive NN-correlations described in \cite{Alvioli:2009ab}.

\begin{figure*}[ht!]
  \centering
  \includegraphics[width=0.8\linewidth]{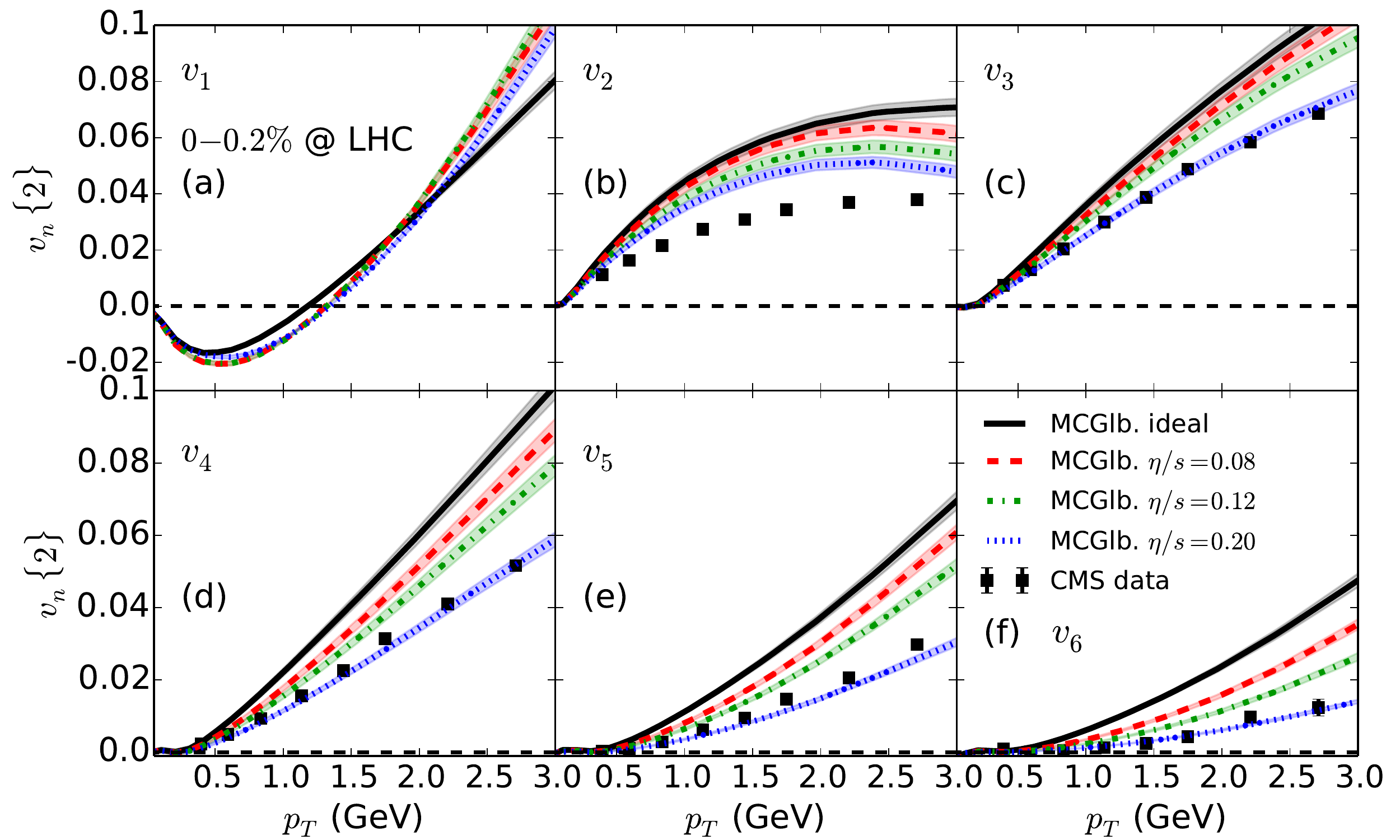}
  \caption{(Color online) MC-Glauber model calculations, including realistic NN correlations in the initial 
  state, of the $p_T$-differential two-particle cumulant $v_n\{2\}$ ($n \in [1, 6]$) of charged hadrons, 
  compared with CMS measurements for 2.76\,$A$\,TeV Pb+Pb collisions at 0-0.2\% centrality 
  \cite{CMS:2013bza}. }
  \label{fig5}
\end{figure*}

Since the initial eccentricities are small on average ($\varepsilon_n\{2\}{\,\sim\,}0.07{-}0.12$ for MC-KLN and $\varepsilon_n\{2\}{\,\sim\,}0.12{-}0.17$ for the MC-Glauber model), non-linear mode couplings \cite{Qiu:2011iv} in the higher order ($n{\,>\,}3$) anisotropic flow coefficients $v_n$, involving the product of multiple $\varepsilon_n$ \cite{Gardim:2011xv,Teaney:2012ke}, are suppressed. We expect a dominantly linear response between the initial $\varepsilon_n\{2\}$ and the final $v_n\{2\}$.\footnote{%
   A recent analysis \cite{Mazeliauskas:2015vea} shows that even in central collisions there is
   a subleading contribution to the triangular flow $v_3$ that is driven by a different radial moment
   $\varepsilon_{3,5}$ of the initial triangularity. Still, the response to the linear superposition of
   these two dominant drivers is linear \cite{Mazeliauskas:2015vea}.}
The conversion efficiency $v_n\{2\}/\varepsilon_n\{2\}$ is controlled by the specific shear viscosity of the medium, as seen in Fig.~\ref{fig4}. Whereas all simulations capture the general trend of increasingly strong suppression of $v_n$ for higher harmonics $n$, which is caused by the non-zero shear viscosity of the medium, both initial condition models fail to quantitatively reproduce the measured $v_n\{2\}$ spectrum ($v_n\{2\}$ as a function of $n$). This statement holds for all the $\eta/s$ values explored in the simulations. 

In the CMS data, the magnitude of $v_2\{2\}$ is very close to $v_3\{2\}$, which can not be reproduced, neither with MC-Glauber nor with MC-KLN initial conditions.\footnote{%
   The authors of Refs.~\cite{Denicol:2014ywa,Rose:2014fba} argue that accounting for 
   short-range correlations between nucleons in the Monte Carlo sampling of the nucleon
   positions within the colliding nuclei and for bulk viscous effects somewhat alleviates this 
   problem. However, the experimentally observed rapid change of the ratio 
   $v_2\{2\}$/$v_3\{2\}$ in ultra-central multiplicity bins of increasing width, from a value
   ${\approx\,}1.0$ at 0-0.2\% centrality to ${\approx\,}1.15$ at 0-2.5\% centrality, rising to
   almost 1.5 at 2.5-5\% centrality \cite{CMS:2013bza}, remains unexplained.}
In hydrodynamic simulations, the conversion efficiency from initial $\varepsilon_n\{2\}$ to final $v_n\{2\}$ is expected to decrease with increasing order $n$ \cite{Alver:2010dn,Schenke:2011bn}, roughly as $\ln[v_n\{2\}/\varepsilon_n\{2\}]{\,\propto\,}{-}\frac{\eta}{s}n^2$ \cite{Lacey:2013qua}. Given this expectation and the hydrodynamic results shown in Fig.~9 of Ref.~\cite{Schenke:2011bn} and Fig.~\ref{fig4} here, an observed ratio of $v_2\{2\}/v_3\{2\}{\,\approx\,}1$ in ultra-central Pb+Pb events would require a ratio of the corresponding initial eccentricities $\varepsilon_2\{2\}/\varepsilon_3\{2\}{\,\sim\,}0.5{-}0.7$, depending on $\eta/s$. This is clearly inconsistent with the predictions from the MC-Glauber and MC-KLN models shown in Fig.~\ref{fig2}b. Starting with an initial fluctuation spectrum featuring $\varepsilon_2\{2\}{\,\simeq\,}\varepsilon_3\{2\}$, as shown in Fig. \ref{fig2}b, for reasonable values of $\eta/s$ the final $v_2\{2\}$ from hydrodynamic simulations will always be about 30\% larger than $v_3\{2\}$. Any theoretical calculation using such initial conditions will therefore overestimate the measured $v_2/v_3$ ratio (at least as long as only shear viscous effect are included). Within the current hydrodynamic framework, to reproduce a $v_n$ hierarchy similar to the one observed by CMS, we need an initial condition model that provides a triangular deformation $\varepsilon_3\{2\}$ that is significantly larger than $\varepsilon_2\{2\}$ in these ultra-central collisions. 

\begin{figure*}[ht!]
  \centering
  \includegraphics[width=0.8\linewidth]{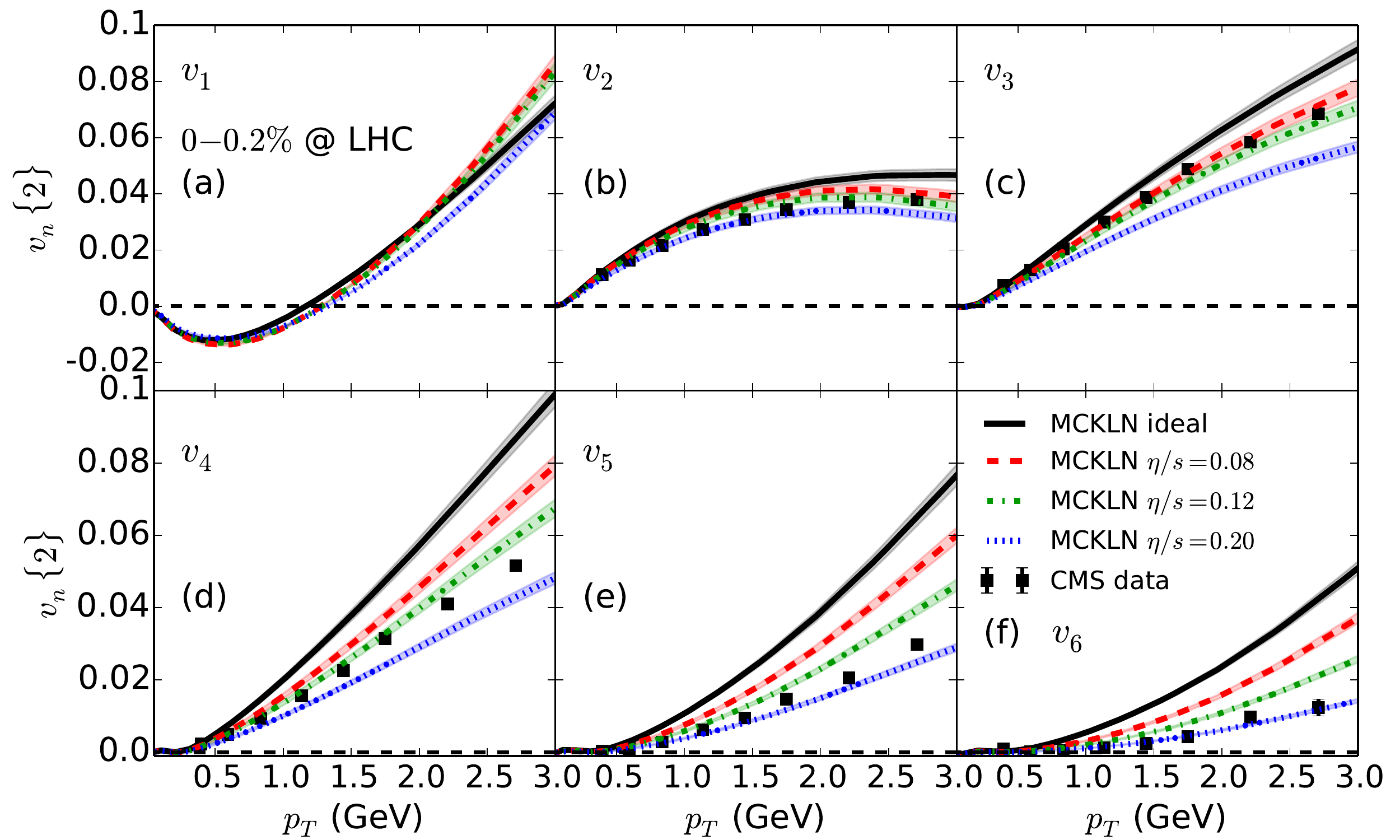}
  \caption{(Color online) Similar to Fig. \ref{fig5},  but for MC-KLN initial conditions. }
  \label{fig5b}
\end{figure*}

Comparing the overall magnitudes of the predicted $v_n\{2\}$ values with the CMS data \cite{CMS:2013bza} in Fig.~\ref{fig4} we find that the data prefer a relatively large specific shear viscosity, for both MC-Glauber ($\eta/s{\,\simeq\,}0.3$) and MC-KLN ($\eta/s{\,\simeq\,}0.2$) initial conditions. In the MC-Glauber case, this conclusion differs from that drawn from earlier studies that did not include $pp$ multiplicity fluctuations. The increased initial r.m.s. eccentricities caused by multiplicity fluctuations generate larger r.m.s. flow anisotropies unless tempered by a larger shear viscosity. Since our MC-KLN initial conditions currently do not yet include the effects of $pp$ multiplicity fluctuations, the $\eta/s$ values required by that model may further increase once this model deficiency is repaired. 

Next, we study the $p_T$-differential anisotropic flows, $v_n(p_T)$. Since for ultra-central collisions the anisotropic flow coefficients are dominated by the fluctuating bumpiness of the initial energy density distributions, the associated flow angles are randomly oriented relative to the (very small) impact parameter and correlated with each other only through the density fluctuations and not by overlap geometry. In such fluctuation-dominated situations, the flow angles $\Psi_n(p_T)$ feature particularly strong $p_T$-dependence, oscillating around their $p_T$-averaged values in patterns that strongly fluctuate from event to event \cite{Heinz:2013bua}. Different flow observables correspond, in general, to different moments of the probability distribution that governs the event-by-event fluctuations of $v_n$ and $\Psi_n$. The specific flow extraction method used in the CMS measurements determines the following $p_T$-differential analog of the two-particle cumulant flow of charged hadrons:
\begin{equation}
\label{eq4}
 v_n\{2\}(p_T) = \frac{\langle v_n(p_T) v^\mathrm{ref}_n 
                          \cos[n(\Psi_n(p_T){-}\Psi^\mathrm{ref}_n)]\rangle}
                          {v^\mathrm{ref}_n\{2\}}.
\end{equation}
Here $v^\mathrm{ref}_n\{2\}$ and $\Psi^\mathrm{ref}_n$ are the $p_T$-{\it integrated} $n$-th order flows obtained from a set of reference particles, taken from a different sub-event (usually a neighboring rapidity bin) to eliminate non-flow and self-correlations. As done in the CMS analysis, we choose all charged hadrons with transverse momenta between 1 and 3 GeV as the reference particles; in the experiment, this choice optimizes the sensitivity for the measurement of higher order $v_n$. Since the theoretically calculated momentum distributions are free of non-flow effects and continuous, corresponding to effectively infinite statistics in a single event, self correlations can be neglected, avoiding the need for different sub-events for signal and reference particles.

In Figs.\,\ref{fig5} and \ref{fig5b} we show the $p_T$-differential flows $v_n\{2\}(p_T)$ at 0-0.2\% centrality from the MC-Glauber and MC-KLN models, compared with the CMS data. We find that the direct flow $v_1\{2\}$ is insensitive to the $\eta/s$ value used in the simulations. It flips from negative to positive at $p_T \sim 1.2$\,GeV due to global momentum conservation. Comparing $v_2\{2\}(p_T)$ through $v_6\{2\}(p_T)$ with the CMS data \cite{CMS:2013bza}, our results using the MC-Glauber model with $\eta/s = 0.20$ provide a fairly good description, except for the  differential elliptic flow $v_2\{2\}(p_T)$ which is overestimated by the calculation. The MC-KLN initial conditions with $\eta/s = 0.20$ can describe $v_2\{2\}(p_T)$ better but underestimate $v_3\{2\}(p_T)$. Although neither model describes all the data equally well, the overall picture seems to favor a QGP shear viscosity in the hydrodynamic simulations that lies at the upper end of the explored range, $(\eta/s)_\mathrm{QGP}{\,\sim\,}0.20$. 

\begin{figure*}[h]
  \centering
  \includegraphics[width=0.9\linewidth]{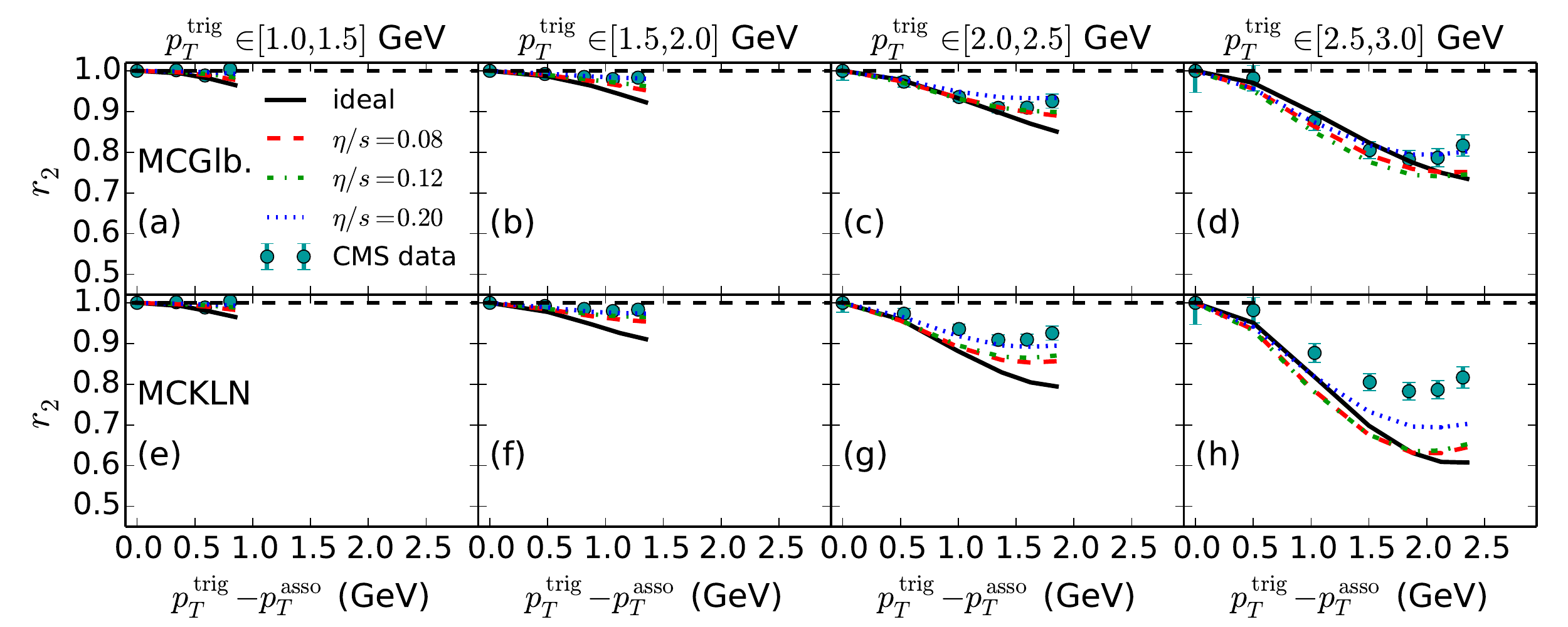}
  \caption{(Color online) Flow factorization ratio $r_{n}(p_T^\mathrm{trig}, p_T^\mathrm{asso})$ ($n = 2, 3, 4$) 
    from model calculations with MC-Glauber (a-d) and MC-KLN  (e-h) initial conditions,
    ultra-central 2.76\,$A$ TeV Pb + Pb collisions at 0-0.2\% centrality. Four values of 
    $\eta/s$ were explored in each case, as indicated. Theoretical results are compared with 
    measurements by the CMS collaboration \cite{CMS:2013bza}.}
  \label{fig6}
\end{figure*}
\begin{figure*}[h]
  \centering
  \includegraphics[width=0.9\linewidth]{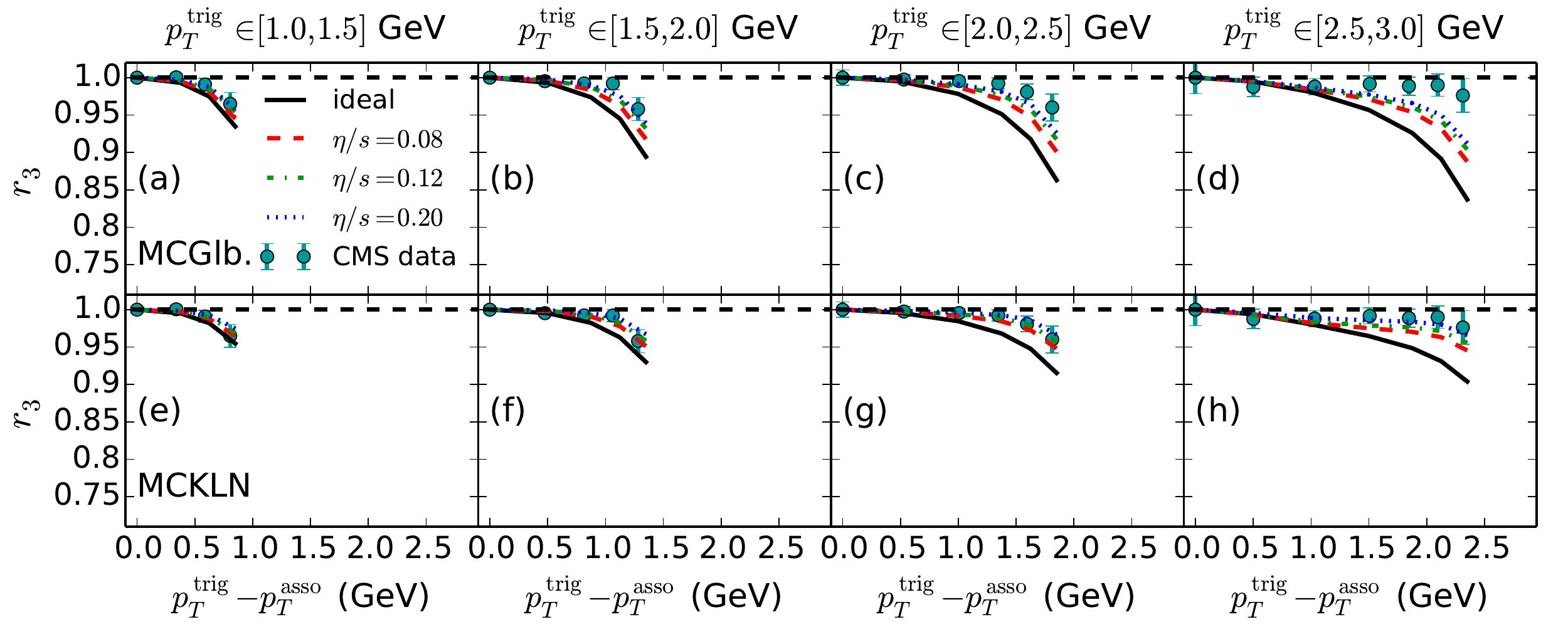}
  \caption{(Color online) Similar to Fig.~\ref{fig6}, but for $n = 3$. }
  \label{fig6b}
\end{figure*}
\begin{figure*}[h]
  \centering
  \includegraphics[width=0.9\linewidth]{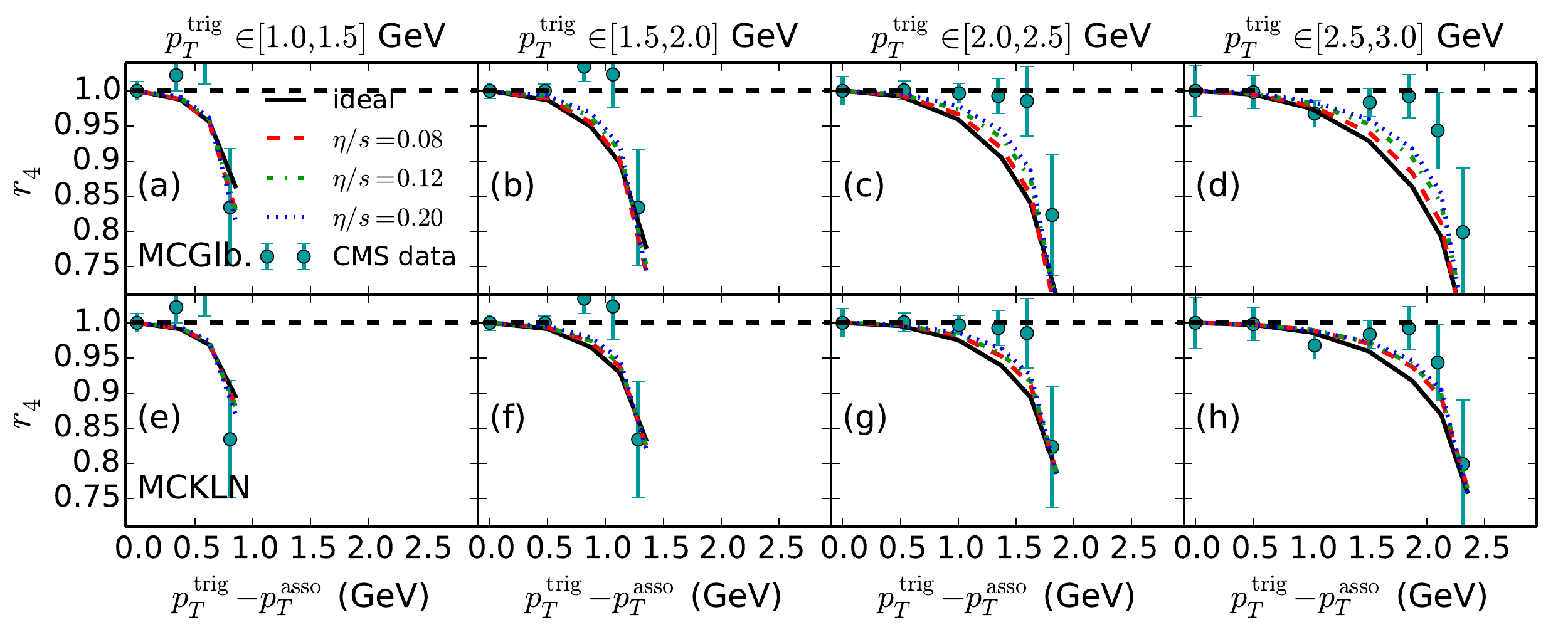}
  \caption{(Color online) Similar to Fig.~\ref{fig6}, but for $n = 4$.}
  \label{fig6c}
\end{figure*}

\section{Flow factorization}
\label{sec4}

In a single event, hydrodynamic flow effects on two-particle correlations factorize into a product of single-particle flow coefficients \cite{Aamodt:2011by,Gardim:2012im}. Due to event-by-event flow fluctuations this factorization is broken in measurements that are based on event averaged observables \cite{Gardim:2012im,Heinz:2013bua}. The factorization breaking effects can be quantified by the ``flow factorization ratios'' 
\begin{eqnarray}
\label{eq5}
  &&r_n(p_\mathrm{T1},p_\mathrm{T2}) := 
  \frac{V_{n\Delta}(p_\mathrm{T1},p_\mathrm{T2})}
         {\sqrt{V_{n\Delta}(p_\mathrm{T1},p_\mathrm{T1})
                  V_{n\Delta}(p_\mathrm{T2},p_\mathrm{T2})}}
 \\\nonumber 
 && = \frac{\langle v_n(p_\mathrm{T1})v_n(p_\mathrm{T2}) 
                 \cos[n(\Psi_n(p_\mathrm{T1}){-}\Psi_n(p_\mathrm{T2}))]\rangle}
                {\sqrt{\la v_n^2(p_\mathrm{T1})\ra\la v_n^2(p_\mathrm{T2})\ra}},
\end{eqnarray}                
where $V_{n\Delta}(p_\mathrm{T1},p_\mathrm{T2})$ are the two-particle differential flow coefficients \cite{Gardim:2012im,Heinz:2013bua} characterizing the correlation between a ``trigger particle" at transverse momentum $p_T^\mathrm{trig}=p_\mathrm{T1}$ and an ``associated particle'' from the same event at $p_T^\mathrm{asso}=p_\mathrm{T2}$. A deviation of this ratio from one indicates that flow factorization is broken. In fact, if $r_n$ is larger than one, factorization must be broken by non-flow effects -- hydrodynamic flow fluctuations always lead to $r_n{\,\leq\,}1$ \cite{Gardim:2012im}.

In Figs.~\ref{fig6} - \ref{fig6c} we show the flow factorization ratios $r_{2,3,4}$ at 0-0.2\% centrality, for different values of $\eta/s$. For the given ranges of trigger particle transverse momentum, these ratios are shown as a function of $p_T^\mathrm{trig}{-}p_T^\mathrm{asso}$ and compared with the CMS measurements \cite{CMS:2013bza}.  

Figure~\ref{fig6} reveals that in ultra-central Pb+Pb collisions the ratio $r_2$ exhibits a large breaking of flow factorization at large $p_T^\mathrm{trig}{-}p_T^\mathrm{asso}$. This breaking is not due to the onset of non-flow effects at large $p_T$ as originally suspected \cite{Aamodt:2011by}, but due to flow fluctuations caused by initial-state fluctuations and very well described by hydrodynamic evolution of the latter. We have checked that about half of the factorization breaking arises from the flow angle fluctuations \cite{Heinz:2013bua} while the other half comes from fluctuations of the magnitudes the anisotropic flows. The figure shows that $\eta/s$ affects the ratio $r_2$ non-monotonically, and we found that this is associated with a change in the relative contribution of $v_n$ and $\Psi_n$ fluctuations to the breaking of flow factorization. This may indicate a non-trivial interplay of viscous damping effects on the fluctuations in flow magnitude and flow angle. Indeed, our studies showed that increasing the shear viscosity suppresses $v_n$ fluctuations but strengthens flow-angle correlations \cite{Qiu:2012uy}.

Comparing the upper panels with the lower ones in Fig.~\ref{fig6}, we find that, irrespective of the choice of $\eta/s$, the ratio $r_2$ deviates from 1 more strongly when we use MC-KLN initial conditions than for MC-Glauber profiles. Similar statements, but with the opposite sign, hold for $r_3$ and $r_4$, shown in Figs.~\ref{fig6b} and \ref{fig6c}: in their case, the factorization breaking effects are weaker for MC-KLN initial conditions than for MC-Glauber ones, again with little sensitivity to $\eta/s$. This implies that $r_n$ actually responds more strongly to changes in the initial fluctuation spectrum than to variations of the shear viscosity. The CMS $v_2$ factorization breaking data in ultra-central collisions, shown in Fig.~\ref{fig6}, favor MC-Glauber-like initial-state fluctuations over the MC-KLN ones, but Figs.~\ref{fig6b} and \ref{fig6c} lead to the opposite conclusion: while the flow factorization breaking effects for triangular ($r_3$) and quadrangular flow ($r_4$) are smaller than for elliptic flow  ($r_2)$, and therefore harder to measure, they seem to be slightly better described by hydrodynamics with MC-KLN initial conditions than for MC-Glauber initial profiles. While the sensitivity to $\eta/s$ is small, $r_3$ and $r_4$ appear to give a slight preference to larger $\eta/s$ values, consistent with the qualitative conclusion drawn in the preceding section from the overall trend of the $v_n$ power spectrum. 

In summary, flow factorization breaking effects appear to open a window on the initial fluctuation spectrum that is only slightly blurred by uncertainties in the shear viscosity during the dynamical evolution from initial to final state. This observable therefore complements the $n$-dependence of the magnitudes of the $p_T$-integrated flow coefficients $v_n\{2\}$ for which shear viscosity effects dominate over variations in the initial-state fluctuation spectrum. While neither the MC-Glauber nor the MC-KLN initial fluctuation spectrum allows to quantitatively reproduce all available data simultaneously, this observation suggests that, by using the full complement of experimentally accessible flow and fluctuation observables, it will in the future be possible to identify the correct initial-state model {\it and, at the same time,} quantify the quark-gluon plasma transport properties.

\section{Conclusions}
\label{sec5}

In this paper, we presented anisotropic flow studies for 0-0.2\% ultra-central Pb+Pb collisions at the LHC, using the MC-Glauber and the MC-KLN initial condition models and evolving the system hydrodynamically with $\eta/s = 0, 0.08, 0.12$, and $0.20$. In the MC-Glauber model, we implement multiplicity fluctuations in the initial state, which boost the initial eccentricities $\varepsilon_n\{2\}$ for $n{\,\leq\,}9$ by 15-45\%. A comparison with CMS data reveals that both MC-Glauber and MC-KLN models fail to reproduce the $p_T$-integrated $v_n$ hierarchy, especially the $v_2\{2\}/v_3\{2\}$ ratio. Further comparisons with the $p_T$-differential $v_n$ tend to favor a relatively large average value of $\eta/s{\,\gtrsim\,}0.2$ for the medium. 

We found a large breaking of flow factorization for the elliptic flow coefficient in ultra-central collisions. Both the fluctuations of the flow magnitude and of the flow angle are important contributors to this breaking. Although our simulations can not fully reproduce the measured $v_n$ spectrum, calculations of the $r_n$ agree overall quite well with the CMS measurements. All qualitative features and trends of the CMS data are correctly reproduced by the hydrodynamic model. Consistent with the conclusion from the $v_n$ comparison, the CMS $r_n$ data again slightly favor $\eta/s{\,\sim\,}0.2$ over smaller values of the average specific shear viscosity during the dynamical evolution.

\acknowledgments{We acknowledge fruitful and stimulating discussion with Wei Li. This work was supported by the U.S. Department of Energy, Office of Science, Office of Nuclear Physics under Awards No. \rm{DE-SC0004286} and (within the framework of the JET Collaboration) \rm{DE-SC0004104}, and in part by the Natural Sciences and Engineering Research Council of Canada.}



\end{document}